\documentclass[11pt,a4paper]{article}
\usepackage[utf8]{inputenc}
\usepackage{amsmath}
\usepackage{amssymb}
\usepackage{amsfonts}
\usepackage[backend=biber, style=numeric, sorting=nty, terseinits=true, giveninits]{biblatex}

\usepackage[hypertexnames=false]{hyperref}
\hypersetup{
    colorlinks=true,
    citecolor = magenta,
    linkcolor=blue,
    filecolor=magenta,      
    urlcolor=blue,
}
\usepackage{caption}
\usepackage{subcaption}
\usepackage{graphicx}
\usepackage{geometry}
\usepackage{hyphenat}
\usepackage{csquotes}
\usepackage[english]{babel}
\usepackage{multirow}
\usepackage{bbm}
\usepackage{algorithm}
\usepackage{algpseudocode}

\defineshorthand{~=}{\hyp{}}
\usepackage{bm}
\usepackage[parfill]{parskip}
\newcommand{\ts}{\textsuperscript}
\usepackage[noblocks]{authblk}

\title{\sc Modelling the Stochastic Importation Dynamics and Establishment of Novel Pathogenic Strains using a General Branching Processes Framework}

\author[1,*]{Jacob Curran-Sebastian}
\author[1]{Frederik Mølkjær Andersen}
\author[1, 2]{Samir Bhatt}

\affil[1]{Section of Epidemiology, Department of Public Health, University of Copenhagen, Copenhagen, Denmark}

\affil[2]{MRC Centre for Global Infectious Disease Analysis, Imperial College London, London,
United Kingdom}

\affil[*]{Corresponding Author: jacob.curran@sund.ku.dk}

\date{}

\begin{document}

\maketitle

\begin{abstract}\noindent{The importation and subsequent establishment of novel pathogenic strains in a population is subject to a large degree of uncertainty due to the stochastic nature of the disease dynamics. Mathematical models need to take this stochasticity in the early phase of an outbreak into account in order to adequately capture the uncertainty in disease forecasts.  We propose a general branching process model of disease spread that includes host-level heterogeneity, and that can be straightforwardly tailored to capture the salient aspects of a particular disease outbreak. We combine this with a model of case importation that occurs via an independent marked Poisson process. We use this framework to investigate the impact of different control strategies, particularly on the time to establishment of an invading, exogenous strain, using parameters taken from the literature for COVID-19 as an example. We also demonstrate how to combine our model with a deterministic approximation, such that longer term projections can be generated that still incorporate the uncertainty from the early growth phase of the epidemic. Our approach produces meaningful short- and medium-term projections of the course of a disease outbreak when model parameters are still uncertain and when stochasticity still has a large effect on the population dynamics. }
\end{abstract}
{\bf Keywords:} Branching Process; Early Disease Dynamics; Importation of Cases; Time-Varying Infectivity. 

\section{Introduction}
Early introductions of a novel pathogenic disease into a population that has not had any previous exposure to infection create a large degree of uncertainty. Firstly, the rate and timing of the earliest introduction can have an impact on the early growth and timing of an outbreak, though the extent of this impact has been questioned in the context of border closures during the COVID-19 pandemic \cite{shiraef2022did}. Secondly, once the disease has entered the population, the internal dynamics of the pathogen spread generate uncertainty due to the inherent randomness of interactions between infectious and susceptible individuals and individual variation in transmissibility. For example, overdispersion in the number of people infected by a single infectious individual has been shown to affect the probability that a transmission chain goes extinct before an outbreak becomes large \cite{lloyd2005superspreading, kucharski2015role}. Adding to this, for many pathogens, the course of an outbreak depends on individual host characteristics such as time-varying infectivity, and on population-level changes in transmission. 

In these settings, one question of interest is that of when the disease has become established in the population, at which time the stochastic effects of the disease dynamics are essentially negligible. This question has been considered much in the literature on stochastic and deterministic models of infectious diseases \cite{barbour1975duration, diekmann2013mathematical}, but here we restrict attention to the time at which stochastic elimination of the disease within a population is impossible, and thus elimination is only achievable through interventions aimed at reducing transmission. From a modelling point of view, we may also interpret this time as the time at which the deterministic, mean-field model closely approximates the stochastic model. 

For the early stochastic phase of the outbreak, we use a Crump-Mode-Jagers (CMJ) branching process (sometimes referred to as a \emph{general} branching process) \cite{crump1969general, crump1968general, jagers1975branching}. CMJ processes are flexible branching processes, in that they allow infectious individuals to transmit to others at the points of a point process that is allowed to be dependent on time as well as on the \emph{age} of an individual, which in our context corresponds to the time-since-infection of an individual. They also permit individuals to have arbitrary infectious period distributions, allowing for greater variability than is granted by Markov processes (such as, for example, the continuous-time Galton-Watson process, which approximates the standard SIR model in the early, exponential growth phase). CMJ processes were shown by Ball and Donnelly \cite{ball1995strong} to approximate a general model for disease transmission in a closed, well-mixing population up until the number of cases reaches approximately $\mathcal{O}(\sqrt{N})$, where $N$ is the initial number of susceptible people in the population. This is consistent with earlier results on the limit at which branching process models poorly approximate the non-linear disease dynamics, due to Mollison \cite{mollison1977spatial}. These processes have been used recently as a means of modelling infection incidence and prevalence in a unified manner \cite{pakkanen2023unifying, penn2023intrinsic}, as well as estimating the time of first detection of a case in the context of COVID-19 \cite{czuppon2021stochastic}.

We also propose a method of switching between two models, from the stochastic CMJ process to a deterministic model, by calculating the distribution of times taken for a disease to become established in the population via a stochastic model, and then modelling the subsequent deterministic dynamics using integral equations together with a random time shift. This approach was first identified by Metz \cite{metz1978epidemic}, and was demonstrated to converge to the stochastic model of the full epidemic by Diekmann \cite{diekmann1977limiting}. Such hybrid models that incorporate both stochastic and deterministic elements have been considered by a number of authors \cite{binder2012hybrid, rebuli2017hybrid, yan2018distribution}. 

We further extend this model of disease transmission by allowing cases to be imported from an exogenous source according to an independent Poisson process, such that each importation event gives rise to an independent CMJ process. This results in a more realistic model of early transmission, which is rarely only seeded by a single case but, rather, is seeded by multiple importations over time such that cases are replenished even when local transmission chains die out. These, and other types of branching processes, including multitype and Bellman-Harris processes, have a wide number of applications in biology beyond infectious diseases \cite{alexandersson2000application, haccou2005branching, kimmel1992branching, sindi2013discrete}, as do continuous-state diffusion approximations of these processes \cite{feller1951diffusion, jagers1975branching}.

\section{Methods}

\subsection{Internal Disease Dynamics} \label{sec1_IDD}
We model the outbreak of a novel strain in a population that is totally susceptible using a general Crump-Mode-Jagers (CMJ) branching process as an approximation for the early, exponential growth phase of the epidemic, when depletion of the susceptible population has a negligible impact on the disease dynamics \cite{crump1968general, jagers1975branching}. We further assume that all individuals in the population mix homogeneously and act independently of one another.

Let $Z(t, v)$ denote the \emph{prevalence} of infection at time $t$, i.e. the number of individuals infectious at time $t$, starting with a single infectious individual at time $t=v$. In practice, we will set $v = 0$ and drop $v$ from the notation, referring to the prevalence as $Z(t)$. Suppose, in addition, that the infectiousness of an individual depends on the time since they were initially infected (often referred to as an individual's \emph{age} in the branching process literature), which we denote $\tau$, and that individuals recover after a random time $\mathcal{L}$. We model the disease spread using a Crump-Mode-Jagers process, which is determined by the following ingredients:
\begin{enumerate}
    \item $L(\tau, v)$ - The cumulative distribution function (CDF) of the infectious period $\mathcal{L}$ for all individuals, which is the probability that an individual is infectious up until time-since-infection $\tau$ and until time $v$. We assume that this function is differentiable and denote its derivative (i.e. the probability density function, or PDF, of the infectious period) by $l(\tau, v)$.  
    \item $K(t, \tau)$ - A reproduction process, which is a counting process that determines the number of secondary infectious cases that an individual produces over their lifetime. We will henceforth assume that $K(\cdot)$ is a Poisson process with rate $\xi(t, \tau)$, though other choices are possible. 
    \item $\chi(\cdot)$ - A random characteristic function that determines how the process is counted. In our analysis, we will restrict our attention to counting the \emph{prevalence} of infection in the population, which is counted using the random characteristic $\chi(u) = 1 \text{ if } u \in [0, \mathcal{L})$ and $\chi(u) = 0$ otherwise. Cumulative incidence can be obtained similarly via a characteristic function, see \cite{pakkanen2023unifying}.
\end{enumerate}
The infectious period distribution, $L(\tau, v)$, which is also referred to in the branching process literature as the lifetime distribution, determines how long individuals remain infectious for once infected. We will mostly not require the infectious period to depend upon time in this paper, and so will drop $v$ from the notation and simply refer to $L(\tau)$. However, in stating some of the results we keep the full time-dependence to show that these are also valid when the infectious period distribution depends on time. In the special case where the infectious period is drawn from an exponential distribution and the transmission rate is constant, the process is Markovian. However, in general, CMJ processes are non-Markovian. 

Choosing the reproduction process to be a (time-inhomogeneous) Poisson process means that infectious individuals may infect susceptible individuals over the course of their infectious period, at a rate that depends on their time since infection, and on time since the start of the outbreak, allowing for more realistic dynamics. We split the intensity of the Poisson process, $\xi(t, \tau)$ following \cite{pakkanen2023unifying} as:
\begin{equation}
    \xi(t, \tau) = \rho(t)k(\tau),
\end{equation}
where $\rho(t)$ is interpreted as a population-level average transmission rate common to all individuals that changes over time. This could be due to innate changes in the disease transmissibility, changes in host behaviour, or interventions introduced at the population level. For example, Ball et al. use a time-varying transmission rate to model an outbreak in a population undergoing a vaccination program such that the proportion of the background population that is immune changes over time \cite{ball2014stochastic}. 

$k(\cdot)$ is an individual rate density that determines the rate at which secondary infectious cases are produced over an infector's lifetime \cite{pakkanen2023unifying}, depending on the individual's time-since-infection. We note that there is a subtle distinction between the change in infectivity of an individuals over their infectious period, and variability in the length of individuals' infectious periods, and it is often the case in time-since-infection models that the parameters $k(\cdot)$ and $(1-L(\cdot))$ are simply combined into a single parameter that determines the variability in infectiousness over an individuals lifetime. For a discussion of this distinction, we refer readers to Chapter 1 of \cite{diekmann2013mathematical}.

The behaviour of the process is governed by the average number of secondary cases produced by a single infectious individual at time $t$, taking into account changes in transmissibility after time $t$. We denote this quantity by $\mathcal{R}(t) = \mathbb{E}[K(t)]$, which is analogous to the \emph{case-reproduction number} defined by Wallinga and Teunis \cite{wallinga2004different}. We have that:

\begin{equation}
    \mathcal{R}(t) := \int_0^\infty \rho(t + \tau)k(\tau) (1-L(\tau)) d\tau. \label{integrated_intensity}
\end{equation}

We will mostly restrict attention to the case where the transmission parameter $\rho(t)$ does not depend on time $t$, so that $\rho(t) = \rho_0$. When $\rho(t)$ is constant, the system is, equivalently, governed by the Malthusian growth parameter, $\alpha$, which is the solution (if a solution exists) to the classical Lotka-Euler equation:

\begin{equation}
1 = \int_0^\infty \mathrm{e}^{-\alpha \tau} \rho_0 k(\tau) (1-L(\tau)) d\tau. \label{malthusian}
\end{equation}

 For $\mathcal{R} > 1$ (where we write $\mathcal{R} = \mathcal{R}(t)$ in the case that $\rho(t)$ is constant), a solution to the above integral always exists. The mean prevalence $\mu(t) = \mathbb{E}[Z(t)]$ grows asymptotically according to the exponential $\mathrm{e}^{\alpha t}$. In the most general case, where the global transmission parameter $\rho(t)$ and the infectious period distribution $L(\tau, v)$ both depend on time, we have the probability generating function (PGF) for the prevalence of the process (see \cite{penn2023intrinsic} for a derivation) as integral equations:
 
\begin{align}
 Q(t, s, v) &:= \mathbb{E}[s^{Z(t, v)}] = s(1 - L(t-v, v)) \mathrm{exp}\left(\int_0^{t-v} [Q(t, s, \tau+v)-1]\rho(\tau + v)k(\tau) d\tau\right) \nonumber \\ &+ \int_0^{t-v}\mathrm{exp}\left(\int_0^\tau [Q(t, s, u + v)-1]\rho(u + v)k(u) du\right)l(\tau, v)d\tau. \label{pgf_prev}
\end{align}

The above is the PGF in the most general case, which accounts for the time-dependent global transmission parameter $\rho(t)$ and lifetime distribution $L(\tau, v)$. In this case, we require the dependence of the PGF $Q(t, u, v)$ on an additional parameter $v$, which denotes the time of the initial case. If, instead, we have $\rho(t) = \rho_0$ and $L(\tau, v) = L(\tau)$ independent of time, the renewal equations for the PGF have a simpler expression without any dependence on the initial condition $v$ \cite{crump1969general}: 

\begin{align}
 Q(t, s) &= s(1 - L(t)) \mathrm{exp}\left(\rho_0 \int_0^t [Q(t-\tau, s)-1]k(\tau) d\tau\right) \nonumber \\ &+ \int_0^t\mathrm{exp}\left(\rho_0 \int_0^\tau [Q(t-u, s)-1]k(u) du\right)l(\tau)d\tau. \label{pgf_prev_simplified}
\end{align}

We may also recover the simple Markovian branching process from the CMJ process by choosing $K(t)$ to be a homogeneous Poisson process, i.e. such that $\xi(\tau) = \beta \tau$ for some $\beta$, and by choosing the recovery period of an individual to be exponentially distributed, i.e. such that $L(\tau) = 1 - \mathrm{e}^{-\gamma \tau}$. In this special case, the continuous-time Galton-Watson process is recovered, and equation \eqref{pgf_prev_simplified} may be written in the form of the usual Kolmogorov backward equation (see, for example, Equation 7.1 on p. 104 of \cite{harris1963theory}):

\begin{equation*}
    \frac{\partial Q(t, s)}{\partial t} = \gamma - (\beta + \gamma)Q + \beta Q^2.
\end{equation*} 

In addition to allowing an individual's infectiousness to vary over the course of their infectious period, one could also extend this approach to make the infectiousness of each individual random, such that each individual exhibits a random infectiousness at time $\tau$ after being infected. This can be done straightforwardly by choosing the counting process $K(t, \tau)$ to be a Cox process \cite{cox1966statistical}, or a doubly-stochastic Poisson process, whose rate parameter $\xi(t, \tau)$ is itself a stochastic process, such that $K(t, \tau)$ conditioned on sample paths of $\xi(t, \tau)$ is a Poisson process. For examples of Cox processes, we refer readers to Chapters 5 and 6 of \cite{pinsky2010introduction}. Another way to incorporate randomness in the infectious profile of individuals would be to extend the CMJ process to a multi-type process, in which individuals are assigned a type corresponding to different (deterministic) infectiousness curves $\xi(t, \tau)$. Each type would then generate cases of different types drawn at random from some set of curves. This may prove a useful way to achieve random heterogeneity in individual infectiousness, particularly when a finite number of infectiousness profiles is desired. We expect that many of our results would generalise straightforwardly to these cases, though we do not consider any explicit examples, and instead make the assumption that all individuals experience the same variable infectiousness at time $\tau$ since infection. 

\subsection{Discretisation of Integral Equation}
In practice we will solely be interested in the process $Z(t) = Z(t, 0)$, i.e. an epidemic starting with a single case at time $v = 0$. Solving the integral equation \eqref{pgf_prev} for the probability generating function $Q(t, s, v)$ with fixed $t$ and $s$ however requires solving $Q(t, s, v)$ for all $v \in [0, t]$. Following \cite{pakkanen2023unifying}, to facilitate the implementation we introduce the auxiliary function 
$Q_c(t, s) := Q(c, s, c-t)$ for $c\geq t \geq 0$. Now evaluating $Q(t, s, 0)$ for a $t\geq 0$ corresponds to solving the equation for $Q_t(t, s) = Q(t, s, 0)$, this function satisfies the integral equation

\begin{align*}
    Q_c(t, s) &= s (1 - L(t, c-t)) \exp\left( \int_0^t Q_c(t-\tau, s) \rho(c - t + \tau) K(d\tau)  \right) \\
    &+ \int_0^t \exp\left( \int_0^\tau Q_c(t - u, s) \rho(c - t + u) K(du) \right) L(d\tau, c - t).
\end{align*}

 Here, the integral equation are further reformulated through Riemann-Stieltjes integrals with $K(\tau) = \int_0^\tau k(u) du$, which provides some numerical stability in the following discretisation procedure. It also allows for the individual infectiousness $k(\cdot)$ and the lifetime distribution function to have discontinuities, which is necessary, for example, if using (discrete) empirical distributions.

We will now approximate $Q_t(t, s)$ over a discrete time-grid $t = 0, \Delta, ..., N\Delta$ for some $N\in \mathbb N$ and $\Delta >0$, through a recursive approximation of $Q_{n\Delta}(i\Delta, s)$ for $i\leq n\leq N$ by right Riemann-Stieltjes sums 

\begin{align}
    \widetilde{Q}_{n\Delta}(i\Delta, s) &:= s (1 - L(i\Delta, (n-i)\Delta)) \exp\left( \sum_{k = 0}^{i-1} \left[\widetilde{Q}_{n\Delta}(k\Delta) - 1\right]\rho\big((n-k)\Delta\big) \Delta K_{i-k} \right) \nonumber \\
    &+ \sum_{j = 0}^{i-1} \exp\left( \sum_{k = j+1}^{i-1} \left[\widetilde{Q}_{n\Delta}(k\Delta)- 1\right] \rho\big((n-k)\Delta\big) \Delta K_{i-k} \right) \Delta L_{i-j}^{(n-i)\Delta}.
    \label{eq:discreteIntEq}
\end{align}

for $i = 1, ..., n$ with

\begin{align*}
    &\Delta K_k := K\big(k\Delta\big) - K\big((k-1)\Delta\big) \\
    &\Delta L_{j}^{v} := L\big(j\Delta, v\big) - L\big((j-1)\Delta, v\big).
\end{align*}

A naive implementation of equation \eqref{eq:discreteIntEq} would require three nested for-loops, which in an interpreted language will make computations inefficient in interpreted languages such as Python or R . We here present an alternative, vectorised implementation. For this, we store the results and intermediary computations in a number of matrices which we present here before the algorithm. We store the values of $\widetilde{Q}_{n\Delta}(i\Delta)$ in a $(N + 1) \times (N+1)$ lower diagonal matrix $\mathbf{Q}$, such that $\mathbf{Q}[n, i] = \widetilde{Q}_{n\Delta}(i\Delta)$, i.e.

\begin{align*}
  \mathbf{Q} : = \begin{bmatrix} 
  \widetilde{Q}_0(0) & 0 & 0 & \cdots & 0 \\
  \widetilde{Q}_{\Delta}(0) & \widetilde{Q}_{\Delta}(\Delta) & 0 & \cdots & 0 \\
  \widetilde{Q}_{2\Delta}(0) & \widetilde{Q}_{2\Delta}(\Delta) & \widetilde{Q}_{2\Delta}(2\Delta) & \cdots & 0 \\
  \vdots & \vdots & \vdots & \ddots & \vdots \\
  \widetilde{Q}_{N\Delta}(0) & \widetilde{Q}_{N\Delta}(\Delta) & \widetilde{Q}_{N\Delta}(2\Delta) & \cdots & \widetilde{Q}_{N\Delta}(N\Delta)
  \end{bmatrix} 
\end{align*}

The first column of the matrix is populated with the known value $Q_t(0) = s$ in all rows. We then compute along each row from left to right until hitting the diagonal, that is, we compute the value of $\widetilde{Q}[n, i]$ from $\widetilde{Q}[n, i-1], \widetilde{Q}[n, i-2], ..., \widetilde{Q}[n, 0]$ for $i \leq n$, the remaining elements for each evaluation can be precomputed and stored appropriately in matrices. We let $\mathbf{L}$ be a $(N+1)\times (N+1)$ lower diagonal matrix such that $\mathbf L[n, i] = L(i\Delta, (n-i)\Delta)$ for $i\leq n\leq N$. Furthermore, let $\mathbf{D_K}$ and $\mathbf{D_L}$ be $N\times N$ upper diagonal matrices such that for $n\leq i \leq N$,

\begin{align*}
    &\mathbf{D_K}[n, i] = \rho\big((N-i+n)\Delta\big)\Delta K_{N-i} \\
    &\mathbf{D_L}[n, i] = \Delta L_{N-i}^{n\Delta}.
\end{align*}

If not specified otherwise, all operations and functions in Algorithm \ref{alg:vec} are applied elementwise, in particular we introduce the elementwise (Hadamard) product $\odot$ of two matrices of the same dimensions. For matrices $\mathbf A, \mathbf B$ with the same number of rows, let $[\mathbf A ; \mathbf B]$ be their column concatenation and let \texttt{ColFlip}, \texttt{RowSum} and \texttt{RowCumSum} be matrix transformations that are respectively reversing the order of the columns in the matrix, summing over the elements of each row and making a cumulative sum over each row. Note that \texttt{ColFlip} and \texttt{RowCumSum} returns matrices of the same dimensions as their input, while \texttt{RowSum} returns a vector with length equal to the number of rows in the input matrix.

\begin{algorithm} 
  \caption{Vectorized implementation}\label{alg:vec}
\begin{algorithmic}[1]
\Require matrices $\mathbf L, \mathbf{D_K}, \mathbf{D_L}$
\Require number of time steps $N$
\Require step size $\Delta$

\State $\mathbf Q \gets \mathbf 0_{(N+1)\times (N+1)}$
\State $\mathbf Q[0:N, 0] = s$

\For{i=1,...,N}
  \State $\mathbf B \gets \big(\mathbf Q[i:N, 0:(i-1)] - 1\big) \odot \mathbf{D_K}[0:(N-i), (N-i):(N-1)]$
  \item[]
  \State $\mathbf B_1 \gets \texttt{ColFlip}\big( \texttt{RowCumSum}\big( \texttt{ColFlip}(\mathbf B) \big) \big)$
  \State $\mathbf{I_1} \gets s \big(1-\mathbf L[i:N, i]\big) \odot \exp(\mathbf B_1[0:(N-i), 0])$
  \item[]
  \State $\mathbf{B_2} \gets \big[ \mathbf B_1[0:(N-i), 1:(i-1)] \, ; \, \mathbf 0_{(N-i+1)\times 1} \big]$
  \State $\mathbf{I_2} \gets \texttt{RowSum}\big( \exp(\mathbf B_2) \odot \mathbf{D_L}[0:(N-i), (N-i):(N-1)] \big)$
  \item[]
  \State $\mathbf Q[i:(N+1), i] \gets \mathbf{I_1} + \mathbf{I_2}$
\EndFor
\State \Return $\mathrm{diag}(\mathbf Q)$
\end{algorithmic}
\end{algorithm}

To further improve efficiency in a situation with light-tailed distributions $L$ and $K$, we can, with only a small error in the final approximation, truncate their densities at a finite value making the matrices $\mathbf L, \mathbf{D_K}$ and $\mathbf{D_L}$ sparser. The overall complexity of Algorithm 1 is $\mathcal{O}(n^3)$ but with truncation introducing sparsity (truncation level $m$), the complexity becomes $\mathcal{O}(n^2 m)$, which for long time series ($m\ll n$) is approximately quadratic.  

\subsection{Distribution of Prevalence over Time} \label{Section_PMF}
The probability mass function for the prevalence at time $t$ can be obtained from the generating function \eqref{pgf_prev} via a fast Fourier transform (following \cite{penn2023intrinsic}). The probability generating function is a power series of the form $Q(t, s) = \sum_{k=0}^\infty p_k(t) s^k$, where $p_k(t) = Pr(Z(t) = k)$ is the probability that the prevalence at time $t$ is equal to $k$. For each time $t$, the $k\ts{th}$ Fourier coefficient, $p_k(t)$ can be approximated via a Riemann sum approximation \cite{miller2018primer}, so that:
\begin{equation}
    p_k(t) = Pr(Z(t) = k) \approx \frac{1}{M}\sum_{j=0}^{M-1} Q\left(t, \mathrm{e}^\frac{2 \pi i j}{M}\right)\mathrm{e}^\frac{-2 \pi i kj}{M}. \label{Riemann_sum}
\end{equation}

At each time point $t$, the full distribution of the prevalence can be recovered by first solving the integral equation \eqref{pgf_prev} numerically and then applying the fast Fourier transform in order to recover the probabilities $p_k(t)$ in \eqref{Riemann_sum}. This requires that a sufficiently large value of $M$ be chosen, such that the tail of the distribution is well captured. This approach therefore works best when the branching process is critical or subcritical (i.e. when the Malthusian parameter $\alpha \leq 0$), or in a supercritical process for times $t$ when the number of cases is still small. For large outbreaks, the number of cases explodes quickly, and hence the computational time required to calculate $p_k(t)$ for large $k$ becomes prohibitively large. In our implementation, we parallelise the computation of each term in \eqref{Riemann_sum}, since each term may be calculated independently, thus reducing the time taken to compute the overall sum. 

\subsection{Importation Dynamics}
Given the disease dynamics within a population, we now turn to modelling the importation of infectious individuals into that population via an independent Poisson process \cite{jagers1975branching}, $\Lambda(t)$, which may be inhomogeneous in time. We will, however, restrict ourselves to the case where the Poisson process governing person-to-person transmission does not depend on global time $t$, such that $\rho(t) = \rho_0$. This allows us to use equation \eqref{pgf_prev_simplified} rather than \eqref{pgf_prev} for the generating function of the process, $Q(t,s)$.

Let $\lambda(t)$ be the time-varying rate at which infectious individuals enter the population and suppose that at each importation event a single infectious individual enters the population. $Y(t)$ denotes the process that starts with no cases at time $t=0$, but which allows importation of cases according to the Poisson process $\Lambda(t)$, each of which starts an independent general branching process, counted by $Z(t)$ (assuming a characteristic $\chi$ has been chosen, here we will choose to count the prevalence as before). The CMJ process with importation is thus defined as a \emph{marked} Poisson process, in which each point of the Poisson process that governs importation is an independent CMJ branching process. Let $\theta(n) = \mathrm{inf}\{t : \Lambda(t) \geq n\}$ denote the time at which the number of introductions from the Poisson process reaches $n$. The CMJ process with importation is then defined by:
\begin{equation}
    Y(t) = \sum_{n=0}^{\Lambda(t)} Z(t - \theta(n)), \label{Campbell_sum} 
\end{equation}
which is to say, the process is defined at time $t$ by $\Lambda(t)$ independent branching processes, each starting with a single case at start time $\theta(n)$.

Starting with a population which has no infectious individuals at time $t=0$, we may then write the probability generating function, $R(t, s)$, for the process $Y(t)$ with importation as:
\begin{equation}
    R(t, s) := \mathbb{E}[s^{Y(t)}] = \mathrm{exp}\left(\int_0^t(Q(t-\tau, s) - 1)\lambda(\tau)d\tau\right). \label{R_PGF}
\end{equation}

The above formulation is a consequence of Campbell's Theorem, which deals with random sums of the form \eqref{Campbell_sum} (see \cite{dorman2004garden} for an introduction to marked Poisson processes and Campbell's Theorem applied to branching processes with importation). We may solve Equation \eqref{R_PGF} numerically together with Equation \eqref{pgf_prev} to obtain the PGF of the branching process with importation. We may, further, assume that the importation process $\Lambda(t)$ forms a compound Poisson process \cite{daley2007introduction}, such that the number of arrivals at each event is an independent and identically distributed random variable $X_i$, i.e.
\begin{equation}
    \Lambda(t) = \sum_{i=0}^{\Xi(t)}X_i,
\end{equation}
where $\Xi(t)$ is a (potentially inhomogeneous) Poisson random variable with rate $\lambda(t)$. As an example, if each of the $X_i$ are i.i.d. and follow a Logarithmic Series distribution with probability $p$ (i.e. $X_i \sim \mathrm{Log}(p)$ such that $P(X_i = k) = \frac{-1}{\mathrm{log}(1-p)}\frac{p^k}{k}$, and $\lambda(t) = \lambda$ is constant), then the resulting distribution of $\Lambda(t + s) - \Lambda(t)$ for $s > t$ has a Negative Binomial distribution, i.e. $(\Lambda(t + s) - \Lambda(t)) \sim \mathrm{NegBin}(r, p)$, where $r = \frac{-\lambda s}{\mathrm{log}(1-p)}$. 

We may also suppose that $\Lambda(t)$ is a \emph{compound} Poisson process such that the number of arrivals at each event $i$, $X_i$ is a random variable, with probability generating function $h(s)$. In this case, the generating function for the compound process is given by:

\begin{equation}
    R(t, s) = \mathrm{exp}\left(\int_0^t (h(Q(t-\tau, s)) - 1) \lambda(\tau)d\tau\right). 
\end{equation}

Note that the PGF for the non-compound process \eqref{R_PGF} is a special case of the above expression, since the number of arrivals at each event $X_i$ has the generating function $h(s) = s$. 

Finally, suppose we have an outbreak that begins with $Z_0$ cases at time $t=0$, and that importation occurs as an independent Poisson process (either simple or compound) as described above. Since we have the generating functions for the process with and without importation, the PGF of the full process, which we will denote $H(t, s)$, is obtained via:
\begin{equation}
    H(t, s) = R(t, s) \left(Q(t, s)\right)^{Z_0}.
\end{equation}

\subsection{First-Passage Time Distribution}
We next calculate the distribution of random hitting times, $T$, to a given threshold number of cases, which we denote by $Z^*$, such that the disease dynamics are well-approximated by a deterministic model. This time $T$ is called the First-Passage time (FPT) to $Z^*$. The FPT distribution has been considered previously \cite{barbour1975duration, curran2024calculation} as a means of relating stochastic models to their deterministic approximations. To the knowledge of the authors, no analytic results exist for calculating the distribution of First-Passage Times to a fixed number in a general CMJ branching process. We present here a method for estimating the FPT distribution based on the approximation for the PMF of the branching process given in equation \eqref{Riemann_sum}, and initially assume no importation of cases in the population.

We describe how the FPT distribution allows us to switch from a stochastic model to a deterministic model with a random start time, such that the stochasticity from the early phase of the outbreak is incorporated into the deterministic model. Let $Z^*$ be the prevalence at which the subsequent disease dynamics are well-approximated by a deterministic model, and let $T$ be the (random) hitting time for the process to reach $Z^*$. We wish to ensure that, for $t > T$, the stochastic effects of the model have a negligible impact on the epidemic. Namely, we require that the probability that the outbreak reaches zero cases under the branching process is approximately zero for $t>T$ and, also, that the standard deviation divided by the mean number of cases is approximately constant for $t > T$. These two conditions ensure that the outbreak is established in the population, and that the number of infected individuals is growing approximately exponentially over time. 

The extinction probability, $q(t) = Pr(Z(t) = 0)$, of the process at time $t$ can be obtained straightforwardly by solving the integral equation for the generating function via \eqref{pgf_prev} and at $s=0$, i.e. taking $q(t) = Q(t, 0)$. 

Once a threshold $Z^*$ is chosen, the first-passage time to $Z^*$ can be calculated from the probability mass function for the process obtained in Section \ref{Section_PMF}. Let $F_M(t; Z^*) = Pr(Z(t) \leq Z^*)$ be the probability that the prevalence at time $t$ is less than or equal to $Z^*$, where the probability has been calculated according to \eqref{Riemann_sum}. Here, $M$ must be chosen to be sufficiently high such that the full distribution of the number of cases at time $t$ is captured. This can be evaluated by checking that $F_M(t; \infty) \approx 1$. If $U(Z^*)$ is the first-passage time of the branching process to $Z^*$ (where we drop the subscript $M$ for convenience), whose CDF is $F_U(t; Z^*)$, then we have that:

\begin{equation}
    F_U(t; Z^*) = 1 - \frac{F_M(t; Z^*)}{q(t)}, \label{FPT_calculation}
\end{equation}

where we have conditioned on the process not hitting zero to ensure that the process will, eventually, reach $Z^*$ (provided the process is supercritical).

Finally, we may also calculate the first-passage time distribution to a given number of cases for the process $Y(t)$ that includes importation in an analogous way to the process without immigration. Suppose we wish to obtain the first-passage time to some number $Y^*$ of cases, denoted $V(Y^*)$. Since a super-critical process where importation of cases occurs over time does not have zero as an absorbing state and therefore will not truly reach extinction (provided that the rate of importation $\lambda(t) \not \to 0$ as $t \to \infty$), we do not need to condition on non-extinction in order to be assured that the process will reach $Y^*$ cases. Defining $G_M(t; Y^*) = Pr(Y(t) \leq Y^*),$ we have that:

\begin{equation}
    F_V(t, Y^*) = 1 - G_M(t; Y^*).
\end{equation}

\subsection{Time to Extinction}

For subcritical processes, where extinction is certain, we may perform a similar calculation to \eqref{FPT_calculation} in order to obtain the distribution of times taken to hit zero cases. However, in this case we may obtain the probability of there being no cases in the population immediately from the generating function $Q(t, s)$ by evaluating at $s = 0$, such that we do not need to calculate to the Riemann sum approximation \eqref{Riemann_sum}. This allows us to investigate the time to elimination in the case when an intervention is introduced some time after cases have begun to grow in a population. Suppose that we have an epidemic that is growing until some time $t = t_l, l > 0$, at which point an intervention is introduced such that $\mathcal{R}(t) < 1$ for $t > t_l$ and the epidemic begins to decline. By conditioning on non-extinction of the outbreak prior to the intervention being introduced (i.e. on $\{Z(t) \neq 0 : t \leq t_l \}$), we may obtain the time to extinction of the outbreak following an intervention. This gives a distribution of times for which interventions would need to be in place for in order to be certain of elimination of the disease. The probability of extinction after an intervention, $q_l(t, t_l)$, is given by:
\begin{equation}
    q_l(t ; t_l) = \frac{Q(t, 0) - Q(t_l, 0)}{1 - Q(t_l, 0)}, \text{ for } t > t_l.
\end{equation}

We denote the time at which the process becomes extinct by $T_{\textit{ext}}$. For the process $Z(t)$ that does not include immigration, we have that $Z(t) = 0  \iff  T_{\textit{ext}} \leq t$ and, therefore, that the probability $q_l(t ; t_l) = P(T_{\textit{ext}}\le t)$ also gives the CDF of the extinction time. The time to extinction after an intervention at time $t_l$ then has the following PDF, denoted by $f_{T_{\textit{ext}}}(t; t_l)$:

\begin{equation}
    f_{T_{\textit{ext}}}(t; t_l) = \frac{d q_l(t; t_l)}{d t}.
\end{equation}

\subsection{Deterministic Approximation of the Disease Dynamics}

It is well known that deterministic models are appropriate approximations for later phases of an epidemic, when stochastic effects become negligible \cite{andersson2000stochastic, diekmann2013mathematical}. Once the number of cases has become large enough, the disease dynamics can be approximated by a set of partial differential equations going back to Kermack and McKendrick \cite{barbour2013approximating, foutel2022individual, kermack1927contribution}. Once this threshold number of cases has been reached, we may also model the susceptibile population, $S(t)$, which also declined determinisitically as the number of infections in the population increases. Let $S(t), I(t)$ denote the number of susceptible and infectious individuals, respectively, at time $t$ in a population of constant size $N$, and let $i(t)$ denote the number of new infections (i.e. the \emph{incidence}) at time $t$. Then, Barbour and Reinert showed that the CMJ process described in Section \ref{sec1_IDD} converges to:

\begin{gather}
    i(t) =  \frac{S(t)}{N}\int_0^t i(t-\tau)\rho(t)k(\tau)(1-L(\tau)) d\tau \label{MVF_noim} \\
    I(t) = \int_0^t i(t-\tau)(1-L(\tau)) d\tau \\
    \text{subject to } \quad I(t_0) = I_0 \nonumber
\end{gather}
in the limit as the population size $N$ becomes large relative to the number of infectious individuals \cite{barbour2013approximating}.

The deterministic approximation allows us to model the impact that depletion of the susceptible population has on the spread of the disease through the population, which cannot be done for the branching process. Once the number of cases in the branching process has become large, Equation \eqref{MVF_noim} provides a basis for switching to a deterministic model, by letting $I_0 = Z^*$. We may then take the random variable $T$, which denotes the random time at which a novel strain has become established in a population, as a random starting time for \eqref{MVF_noim}. This ensures that the temporal uncertainty from the early stochastic phase of the epidemic can be adequately incorporated into projections made forwards in time. 

We may also straightforwardly include importation into the deterministic approximation, though care should be taken either to ensure that the population size $N$ remains constant in spite of cases being imported, or that the population size is upgraded to become a function of time, $N(t)$. The former can be justified by assuming that, the rate of importation has a negligible impact on the size of the overall population or, alternatively, that the total rate of importation of individuals (regardless of their infection status) is equal to the rate at which individuals are removed from the population. Alternatively, we may choose to assume that the number of imported cases into the population is small relative to the population size, as is often the case, and ignore the effect of importation once a large number of cases is reached. 

If $\lambda(t)$ is the rate parameter of the (inhomogeneous) Poisson process governing the importation of infectious cases into the population, and assuming that non-infectious individuals enter into the population at a rate $\nu(t)$ then we have instead that:

\begin{gather}
    i(t) = \lambda(t) + \frac{S(t)}{N(t)}\int_0^t i(t-\tau)\rho(t)k(\tau)(1-L(\tau)) d\tau \label{MVF_im} \\
    \frac{\partial N(t)}{\partial t} = \lambda(t) + \nu(t) \nonumber \\
    I(t) = \int_0^t i(t-\tau)(1-L(\tau)) d\tau \\
    \text{subject to } I(t_0) = I_0. \nonumber 
\end{gather}

Finally, taking $t_0$ to be a random variable drawn from the FPT distribution to $Z^*$ given in Equation \eqref{FPT_calculation}, we have a deterministic model that includes the temporal variability that arises from the early phase of the epidemic. In practice, we may solve Equations \eqref{MVF_noim} and \eqref{MVF_im} once by taking, for example, the mean of the FPT distribution $T^*$ as our starting time, and then translating the curve forwards or backwards in time within the range of values that the FPT distribution takes. In the same way, we can also translate the FPT distribution forwards in time to obtain the distribution of peak timings for the full epidemic. This approach to including a random temporal shift to the deterministic dynamics from the early stochastic phase was first discussed in \cite{metz1978epidemic}, and the convergence to the deterministic model was demonstrated in \cite{diekmann1977limiting}. This approach has appeared in a number of contexts since, including in \cite{janson2014law}, who demonstrate the validity of this approach for an SIR epidemic on a configuration network with applications to vaccination and control. 

Moving from the stochastic to the deterministic model is important for a few reasons. Whilst the method outlined in Section \ref{Section_PMF} allows us to calculate the full PMF for the prevalence of infection, it only computationally feasible whilst the number of cases remains relatively small. This is because, for each value of $0 \leq k < M$ the integral equation \eqref{pgf_prev} needs to be solved and then the inverse Fourier transform applied to the Fourier coefficients $p_k(t)$. Therefore, we only wish to calculate the full distribution using \eqref{Riemann_sum} for the number of cases whilst the prevalence is low, before switching to alternative approaches. Furthermore, Mollison \cite{mollison1977spatial} demonstrated that, once the number of cases in an outbreak reaches the  $O(\sqrt{N})$, with $N$ being the total population size, the probability that a given newly infected individual encounters an individual who has previously been infected is no longer negligible. This breaks the usual assumption for a branching process that individuals can be treated independently of one another, since each individual's offspring distribution at time $t$ would now depend on the presence of other individuals infected up until time $t$. Hence, choosing $Z^* < \sqrt{N}$ allows us to move from a branching process to a model that incorporates depletion of the susceptible population whilst the branching approximation is still suitable. We note also that Barbour and Utev proved that the branching process is a suitable approximation for the epidemic process (modelled as a Reed-Frost epidemic) until $O(N^{\frac{2}{3}})$ susceptible individuals have been infected, in the sense that the likelihoods of the true and approximating epidemic processes remain close up until this threshold is reached \cite{barbour2004approximating}.

\section{Results}
We apply the CMJ model developed in the previous section to investigate the importation dynamics of a novel strain in a population, given different assumptions about the generation time distribution, the infectious profiles of individual hosts and the rate of importation of the pathogen from external sources. We demonstrate the utility and flexibility of this approach in modelling a number of different scenarios, from the establishment of a novel strain in a population to the effect of lockdowns and other non-pharmaceutical interventions on the extinction of a strain, that can be used to generate a range of projections based on different plausible modelling assumptions.

\subsection{Internal Population Dynamics}
We start with a model of an outbreak that begins with a single infectious case at time $t=0$ in an enclosed population. We assume that the infectious period of an individual follows a gamma distribution with a mean of 4.78 days and a standard deviation of 1.98 days (i.e. $\mathcal{L} \sim \Gamma(6.05, 0.81)$) taken from Challen et al. \cite{challen2022meta}, which was fit to data from the First Few One Hundred (FF100) survey of COVID-19 infections in the United Kingdom. We then take infectiousness of an individual, $k(t, \tau)$ over their lifetime to be equal to the PDF of the lifetime distribution. This ensures that the individual-level intensity integrates to 1, so that we may refer to the proportion of an individual's infectiousness that has been transmitted by time-since-infection $\tau$. 

We then model a number of different scenarios based on the reproduction number $\mathcal{R}(t)$, which corresponds to the integrated force of infection for an individual given in Equation \eqref{integrated_intensity}. For the early phase of the epidemic, we take the basic reproduction number $\mathcal{R}_0$, defined as the number of secondary individuals infected by a single individual in a completely susceptible population, to be 1.5 as a baseline, and assume that the population-level transmission rate $\rho(t) = \mathcal{R}_0 = 1.5$ is constant over time. Over short time scales this is a reasonable assumption, as it may take time for control policies to be implemented, let alone for their effects to have an impact on the disease dynamics. We also test scenarios in which the transmission rate may depend on time, in order to model the impact of control measures introduced once the outbreak has started to grow in size. 

\begin{figure}
    \centering
    \includegraphics[scale = 0.4]{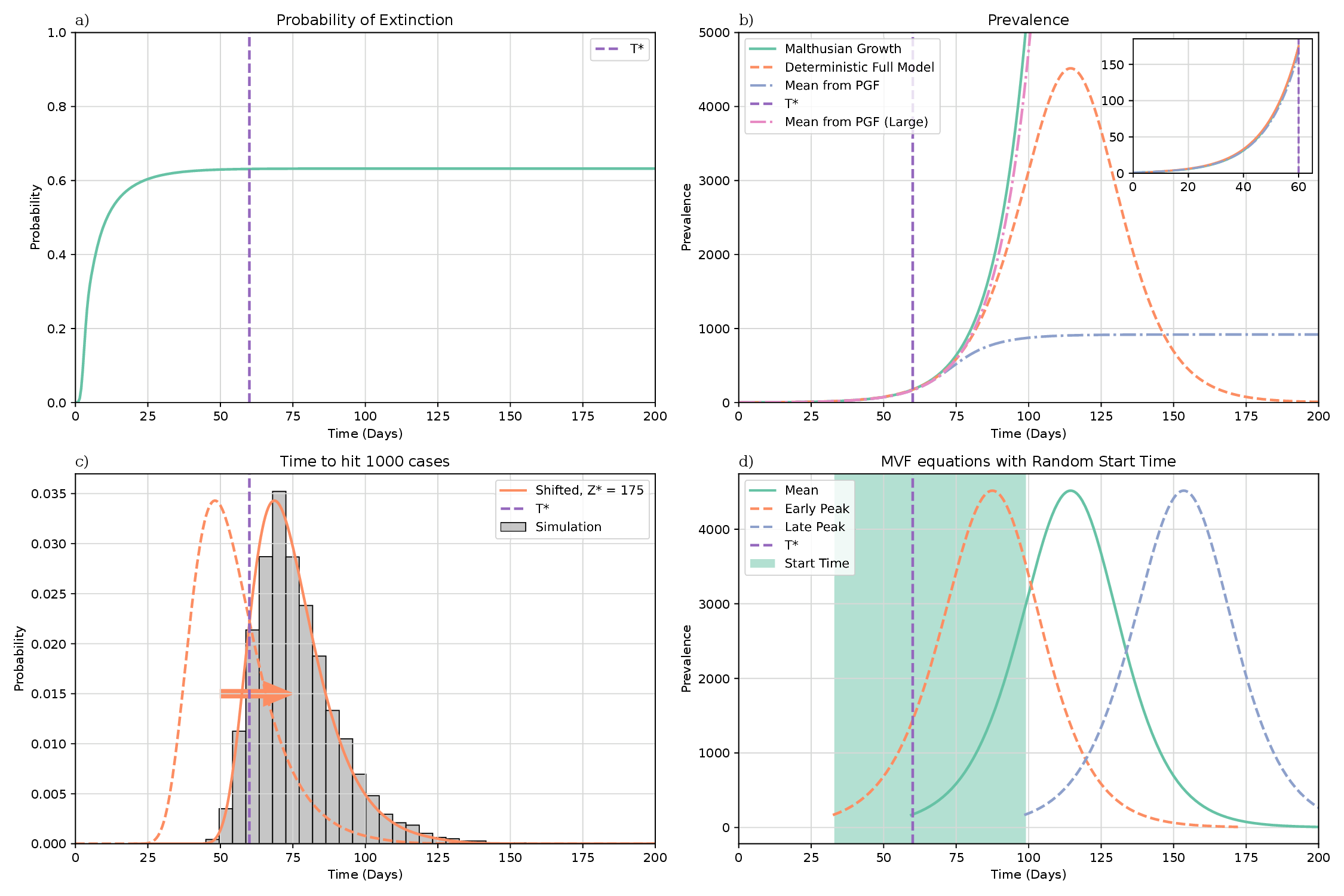}
    \caption{Results from modelling disease dynamics within a population using a Crump-Mode-Jagers process with $R_0=1.5$. a) Cumulative probability of extinction over time for an outbreak that begins with a single case (in the absence of importation). $T^*$ is chosen as the value where this probability becomes approximately constant over time (purple, $T^* = 60$ days). b) Comparison of the mean of the CMJ processes calculated analytically, together with approximations based on Equation \eqref{Riemann_sum} with $M = 10^3$ and $M = 10^5$ (Large). We also show the solution of Equation \eqref{MVF_noim} to show the impact of susceptible depletion. The zoomed plot demonstrates that each of these curves are aligned for $t \leq T^*$. c) The first-passage time distribution for the outbreak to reach 1000 cases, calculated via Equation \eqref{FPT_calculation}, compared with the FPT distribution for $Z^* = 175$ cases shifted forwards in time. d) The resulting outbreak curves calculated from Equation \eqref{MVF_noim} with random start times. The distribution of the peak timing is the same as the distribution of start times shifted forwards deterministically in time.}
    \label{fig:cmj_results}
\end{figure}

Figure \ref{fig:cmj_results} shows the process of moving from the CMJ process to the deterministic approximation via the Kermack-McKendrick equations. We begin by obtaining the probability of extinction of the outbreak over time, $q(t) := P(Z(t) = 0)$, for which we note that, under our baseline assumptions, the ultimate probability of extinction, $q(\infty) = \mathrm{lim}_{t \to \infty} q(t) = 0.63$. We choose our threshold number of cases $Z^*$ based on when $q'(t) < \epsilon = 10^{-3}$, which gives a value of $Z^* = 175$ cases, which is the mean number of cases at time $T^* = 60$ days, i.e. $Z^* = \mathbb{E}[Z(T^*)]$. We also demonstrate in Figure \ref{fig:cmj_results} b) the agreement between the mean of the branching process, calculated both via the Malthusian parameter (Equation \eqref{malthusian}) and the mean of the full distribution of cases (approximated via Equation \eqref{pgf_prev}), and the Kermack-McKendrick equations \eqref{MVF_noim}. This also underscores the importance of choosing $M$ to be large enough in equation \eqref{pgf_prev}, such that the full distribution of the number of cases is captured. Choosing $M = 10^3$ was insufficient for later in the outbreak, when the number of cases becomes large, but remains a good approximation for times before $t = T^*$. 

Panel c) of Figure \ref{fig:cmj_results} demonstrates the suitability of switching from the CMJ process to a deterministic model. Using equation \eqref{FPT_calculation}, we calculate the PDF of the First-Passage Time, $T$, to $Z^* = 175$ cases, as well as to $Z = 1000$ cases. We observe that shifting the FPT distribution forwards in time by $t_{\textit{shift}} = 20.5$ days gives a distribution that is a close approximation to the true FPT distribution. We further confirm that this is a good approximation of the true distribution via stochastic simulation of the process. This motivates switching between the CMJ process and the Kermack-McKendrick equations, the results of which we show in Figure \ref{fig:cmj_results} d), showcasing the difference in peak timings for outbreaks that achieve exponential growth earlier and later than the average.

\subsection{Importation Dynamics and Time to Establishment}
\begin{figure}
    \centering
    \includegraphics[scale = 0.4]{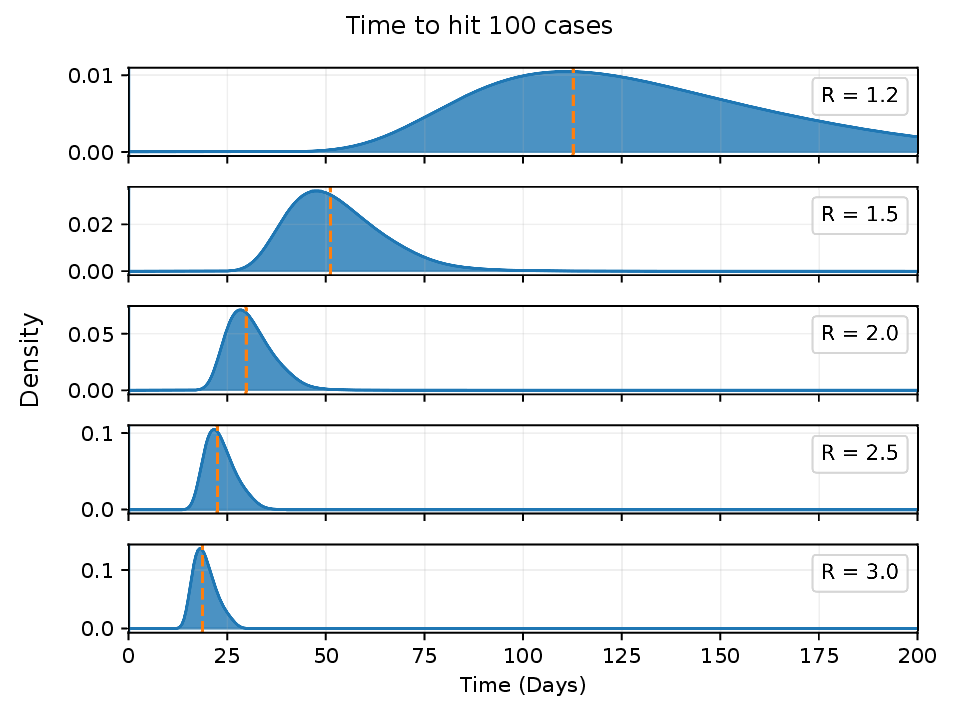}
    \includegraphics[scale = 0.4]{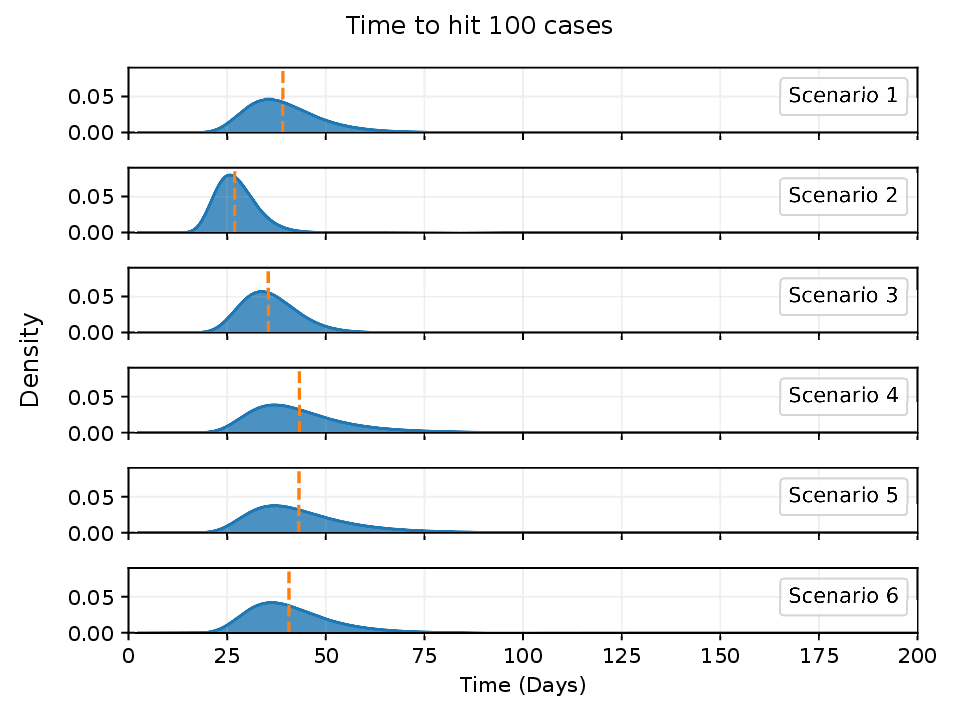} \\
    \includegraphics[scale = 0.5]{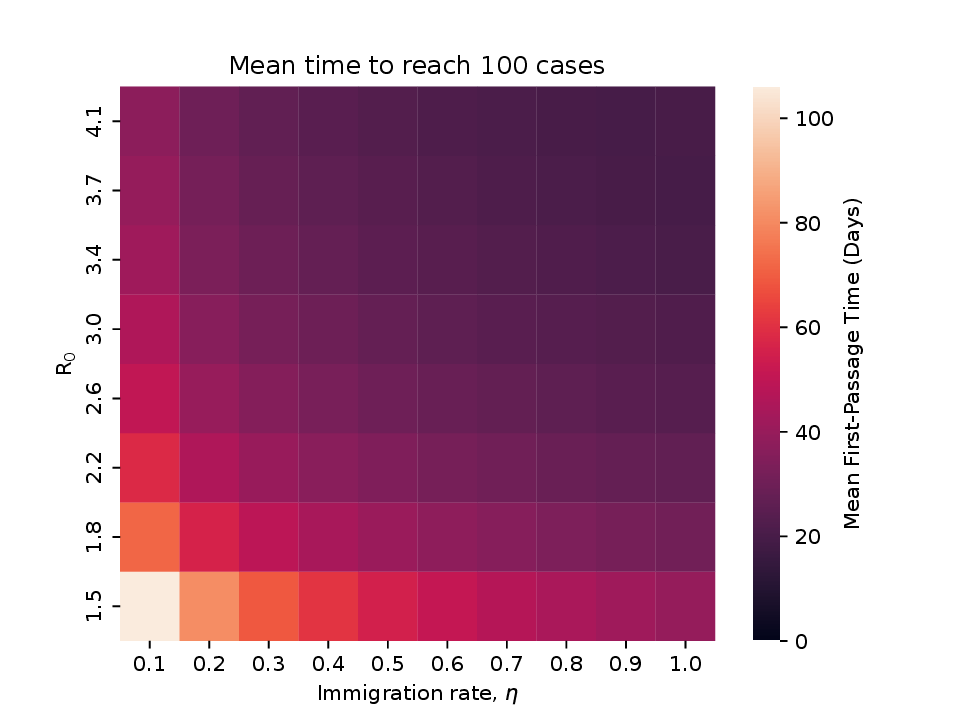}
    \caption{FPT distribution for an outbreak to reach $Z^* = 100$ cases under different control scenarios, obtained using equation \eqref{FPT_calculation}. (Top Left) Each outbreak begins with a single infectious individual, but has different population-level control measures in place, corresponding to different (constant) values of $R$. A higher value of $R$ results in a shorter time to establishment, with lower uncertainty around that time. (Top Right) Each outbreak starts with no cases in the population, but with importation of cases under different scenarios. These scenarios are detailed in Table \ref{tab:im_scenarios}. (Bottom) Mean time for an outbreak to reach 100 cases under different control scenarios aimed at reducing transmission ($\mathcal{R}_0$, kept constant) and importation ($\lambda$, also kept constant).}
    \label{fig:establishment_times}
\end{figure}

Having defined the time to establishment according to \eqref{FPT_calculation}, we now investigate the impact that different control measures have on novel strains becoming established in a population. We consider, firstly, the effect that population-level control measures have on the time to establishment of a strain. These measures correspond in our model to a scaling of the transmission parameter $\rho(t)$ and could take the form of policies such as social distancing, masking or lockdowns. In Figure \ref{fig:establishment_times}, we show the time taken for an outbreak to reach the level $Z^* = 100$ cases for different values of $\mathcal{R}_0$, which we take to be constant over the time period. As expected, the FPT to $Z^*$ decreases monotonically with increasing $\mathcal{R}_0$, as does the variance in the FPT. This demonstrates the impact of introducing measures aimed at reducing $\mathcal{R}(t)$, even if not below the critical threshold of 1, in terms of delaying the growth of an outbreak. 

We also show in the right panel of Figure \ref{fig:establishment_times} the FPT to $Z^*$ cases for a fixed internal $\mathcal{R}_0$ of $1.5$, but with no initial cases and under different importation scenarios. These scenarios are detailed in Table \ref{tab:im_scenarios} and reflect plausible time-dependent importation rates $\lambda(t)$, noting that the importation of cases is generally assumed to be small when compared with the growth of cases due to transmission. Compared with the impact of changing $\mathcal{R}_0$, we see that importation has less of an impact on the FPT distribution. Notably, scenarios 4 and 5 (see Table \ref{tab:im_scenarios}) represent completely successful border policies with a 100\% reduction in imported cases (for scenario 4, this is true in the limit as $t \to \infty$, whereas for scenario 5, imported cases cease to enter the population from outside after 15 days). Both of these scenarios highlight the difficulty in relying on border control policies alone in preventing outbreaks of novel pathogens within a population, a problem that has been noted in \cite{tomba2008simple, hollingsworth2006will}, among others. 

\begin{table}[]
    \centering
    \resizebox{\columnwidth}{!}{%
    \begin{tabular}{|c|c|c|c|}
        \hline Scenario & Description & Functional Form & Parameters \\ \hline
        Scenario 1 & Constant importation at baseline rate & $\lambda(t) = \lambda$ & $\lambda = 0.2$  \\ \hline
        Scenario 2 & Constant importation at a higher rate & $\lambda(t) = \lambda$ & $\lambda = 0.5$  \\ \hline
        Scenario 3 & Exponentially increasing importation & $\lambda(t) = \lambda\mathrm{e}^{\iota t}$ & $\lambda = 0.2, \iota = 0.02$   \\ \hline
        Scenario 4 & Exponentially decreasing importation & $\lambda(t) = \lambda\mathrm{e}^{- \iota t}$ & $\lambda = 0.2, \iota = 0.02$ \\ \hline
        Scenario 5 & Piecewise constant importation & $\lambda(t) = \lambda \mathbbm{1}_{(t < t_l)}$ & $\lambda = 0.2, t_l = 15$ days \\ \hline
        \multirow{2}{*}{Scenario 6}  & \multirow{2}{*}{Two Immigration Sources}  & \multirow{2}{*}{$\lambda(t) = \lambda_1\mathrm{e}^{ \iota_1 t} + \lambda_2\mathrm{e}^{- \iota_2 t}$}   &  $\lambda_1 = 0.05, \iota_1 = -0.02$ \\ & & & $\lambda_1 = 0.01, \iota_1 = 0.2$ \\ \hline

    \end{tabular}
    }
    \caption{importation scenarios corresponding to different (time-dependent) rates for the Poisson process that governs importation of cases. Scenarios 1 and 2 deal with constant importation at different rates. Scenario 3 corresponds to increasing importation of cases at an exponential rate, which occurs when the epidemic external to the population being modeled is growing exponentially. Scenario 4, on the contrary corresponds to an exponentially decreasing importation rate, which models increasing restrictions being placed on individuals entering the population. Finally, scenario 5 models a total shutdown of importation of cases into the population, either through travel bans or screening of cases (in either case, assuming that these measures are 100\% effective).}
    \label{tab:im_scenarios}
\end{table}

In \cite{levesque2021model}, the authors also consider augmenting the counting process $K(t)$ with a binomial distribution that determines whether each infectious individual is contact traced and subsequently isolated, thus shortening their infectious period. This could also be achieved by considering a multi-type CMJ process, which we do not do here. 

\subsection{Impact of Non-Pharmaceutical Interventions} \label{NPIs}
\begin{figure}
    \centering
    \includegraphics[scale = 0.45]{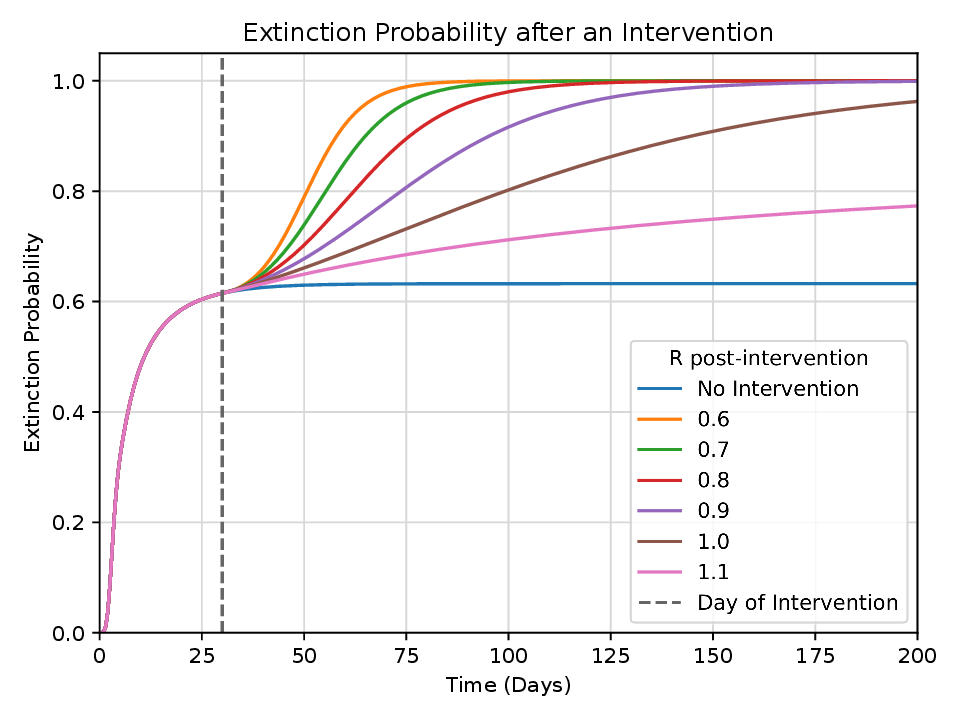}
    \includegraphics[scale = 0.45]{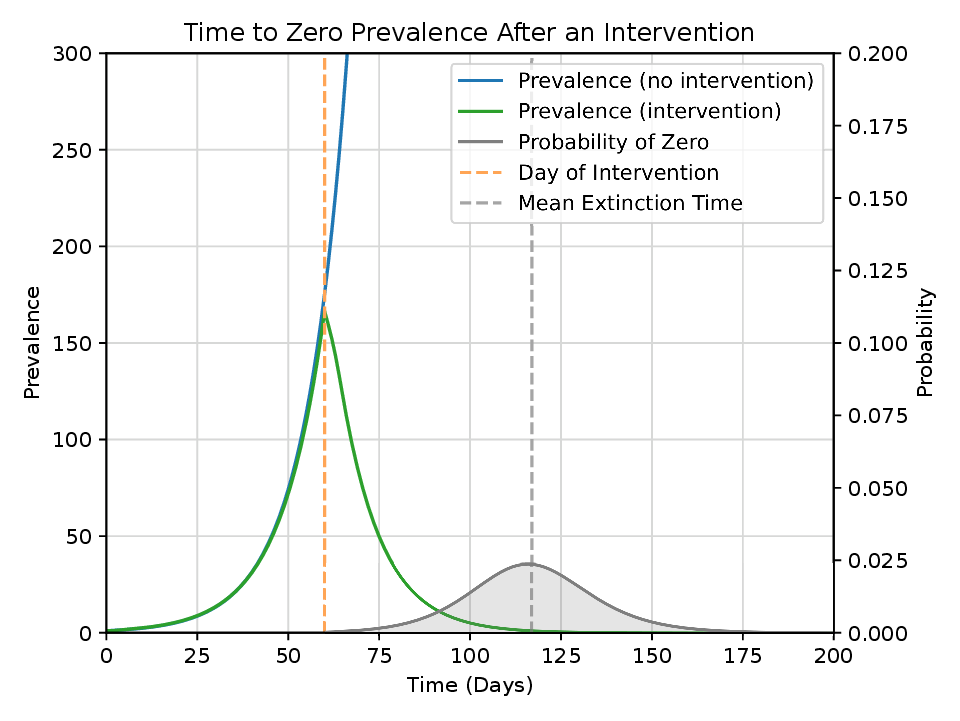} \\
    \includegraphics[scale = 0.5]{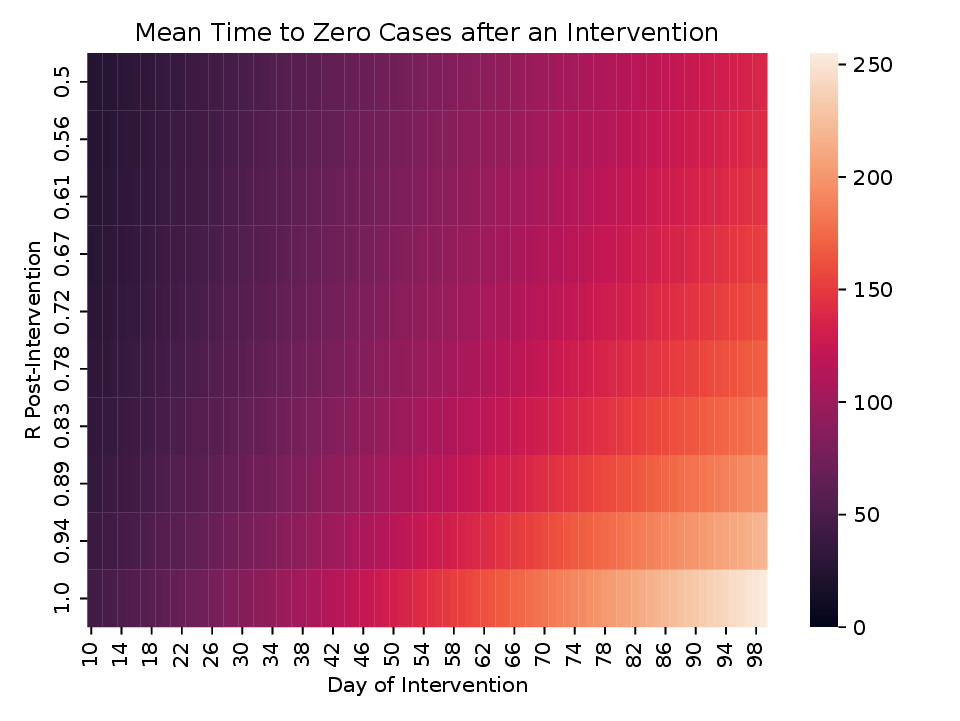}
    \caption{Impact of Non-pharmaceutical interventions aimed at reducing internal transmission via $\mathcal{R}(t)$ (which is kept as a piecewise-constant function, with $\mathcal{R}_0 = 1.5$ pre-intervention), such as a lockdown. In each case, we assume no importation of cases. (Top Left) Cumulative probability of extinction over time for different values of $\mathcal{R}(t)$ post-intervention, which is introduced on day 30. (Top Right) Mean prevalence and distribution of extinction times for an intervention introduced on day 60, reducing $\mathcal{R}(t)$ to 0.75. (Bottom) Mean time to reach zero cases, depending on the value of $\mathcal{R}_0$ post-intervention and on the day that the intervention is introduced.}
    \label{fig:Interventions}
\end{figure}

Finally, we consider the impact that non-pharmaceutical interventions, such as lockdowns, can have on the disease dynamics. We consider a simple model of interventions at the population level, which correspond to an instantaneous shift in the value of $\mathcal{R}(t)$. However, other more complicated forms of $\mathcal{R}(t)$ could be considered to model a gradual change in the reproduction number, by using a sigmoidal function, or periodic changes by using a trigonometric function. For an outbreak starting with a single infectious case and $\mathcal{R}_0 = 1.5$, we consider an instantaneous reduction in the value of $\mathcal{R}(t)$ at different values of $t = t_l$ in Figure \ref{fig:Interventions}. 
Introducing control measures that bring $\mathcal{R}(t)$ below 1 makes the process subcritical, which in turn makes extinction certain. However, the time to extinction varies significantly between scenarios with different values of $\mathcal{R}(t)$ after an intervention. If, in our baseline scenario, measures are introduced on day $t_l = 30$, we see that the probability of extinction is greater than 0.95 on day 64 if $\mathcal{R}(t)$ is brought down to 0.6, compared with day 113 if $\mathcal{R}(t)$ is reduced to 0.9. In addition, we also quantify in Figure \ref{fig:Interventions} the time to reach zero cases after NPIs are introduced on day $t_l = 60$ of an outbreak reducing $\mathcal{R}(t)$ to 0.75. Whilst the mean extinction time occurs at $t = 117$ days, 95 \% of the distribution of extinction times lies between 80 and 156 days. This approach offers insight into how long control measures need to be in place for in order to achieve local elimination of a disease, and can be used for planning interventions with a proper accounting for the stochastic uncertainty that arises in elimination settings. 

\section{Discussion}

We have presented a general model for modelling the early stochastic phase of an outbreak, when the number of cases is relatively small compared to the size of the population in which the outbreak is growing. Our model extends previous models by allowing importation to occur separately via an independent Poisson process, thus capturing a key component of the early phase of an epidemic growing in multiple populations. We define a threshold number of cases at which a novel strain has become established in a population and demonstrate that a deterministic approximation to the full stochastic model is suitable once this threshold has been reached. In the time between the initial case and this threshold being reached, the disease dynamics are highly stochastic, and so we investigate the impact of stochasticity on the establishment of a disease in a population. 

Our approach is centred on solving the integral equations for the probability generating functions \eqref{pgf_prev} and \eqref{R_PGF}. Once these have been solved, the probability mass function can be approximated via \eqref{Riemann_sum} whilst the prevalence in the population is low. This gives the entire distribution of the number of cases for the early phase of the epidemic, where the impact of stochastic effects such as extinction and delayed growth on the overall dynamics is greatest. Furthermore, we use this distribution to calculate the First-Passage Time distribution, which allows us to calculate the time taken for an invading strain to become established in a population. We then demonstrate that this FPT distribution can be used as the random start time for the Kermack-McKendrick equations once the non-linear effects of susceptible depletion become relevant, so that the temporal uncertainty from the early stochastic phase can be incorporated into longer term projections. 

This highlights an important aspect of our approach. The CMJ process gives a full description of the intrinsic aleatoric uncertainty that comes from the early phase of the epidemic. On the one hand, it is computationally cumbersome to solve the integral equations and obtain the PMF for the full stochastic process when cases become large. On the other hand, we demonstrate that the full stochastic process is well approximated by the Kermack-McKendrick model for large numbers of cases, so that computing the PMF for the CMJ process is only required to capture the very early dynamics. 

 Another major advantage of using general Crump-Mode-Jagers processes is their flexibility in modelling aspects of a disease outbreak that are relevant for a particular application of interest. For example, for certain diseases the frequent assumption that the infectious period is exponentially distributed may bias modelling outcomes, whereas it may be reasonable for that same disease to neglect the effects of superspreading. Similarly, different assumptions may be made about the importation of cases from external sources, which we have shown to have an impact on the early growth of an outbreak. CMJ processes allow modellers to keep certain components of the overall model as simple as possible where these do not affect the model outcomes, but change those components that are hypothesised to change the disease dynamics considerably (and indeed to test the impact of these hypotheses). They are also useful for modellers who wish to plan for control measures and interventions under a range of plausible scenarios, which is crucial particularly in the early phase of a pandemic, when precise model parameters and features are less certain. 

 One drawback of using CMJ processes, however, is the difficulty of fitting model parameters to epidemiological data. In \cite{penn2023intrinsic}, the authors developed a likelihood for fitting the full CMJ process to outbreak data, though fitting their full likelihood requires extremely detailed data that is unlikely to be collected during an epidemic, particularly in the early stages. However, they also develop a discrete-time analogue of this likelihood which can be fitted more readily to data, which they demonstrate using the prevalence data from an outbreak of SARS in Hong Kong. The authors of \cite{levesque2021model} also develop a likelihood for fitting CMJ processes to epidemiological data, though with similar drawbacks, which they fit to simulated data. In our case, with the additional inclusion of imported cases, one would also require data on which lineage an infected case belongs to (i.e. knowledge of the initial case in the transmission chain to which each case belongs), which is unfeasible in the majority of cases. In some settings, such as where detailed contact tracing is performed such that infected individuals can be traced back to imported cases, it may be possible to fit our model to data. Instead, we envisage our approach being used primarily as a tool for scenario planning and testing of different hypotheses, rather than a tool for inferring epidemiological parameters, which may be done using other modelling approaches. 

 Finally, we note that, whilst we have chosen to focus on the spread of an infectious disease between individuals in a susceptible population, our modelling approach applies to a much wider range of scenarios in mathematical biology, and could be used for example to model the emergence of antimicrobial resistance in a bacterial population \cite{alexander2012pre, fu2015spatial}, the growth of cancerous cells \cite{avanzini2019cancer, durrett2015branching}, or the evolution of phylogenetic trees \cite{lambert2010contour, lambert2013scaling}. 

\section*{Data Availability}

There are no primary data in the paper; all code and materials required to reproduce the figures in this paper are available at \url{https://github.com/JCurran-Sebastian/CMJ_Branching} and we have archived our code on Zenodo (DOI: 10.5281/zenodo.11108605).

\section*{Acknowledgements}

S.B. acknowledges funding from the MRC Centre for Global Infectious Disease Analysis (reference MR/X020258/1), funded by the UK Medical Research Council (MRC). This UK funded award is carried out in the frame of the Global Health EDCTP3 Joint Undertaking. S.B. acknowledges support from the National Institute for Health and Care Research
(NIHR) via the Health Protection Research Unit in Modelling and Health Economics , which is a partnership between the UK Health Security Agency (UKHSA), Imperial College London, and the London School of Hygiene \&; Tropical Medicine (grant code NIHR200908). (The views expressed are those of the authors and not necessarily those of the UK Department of Health and Social Care, NIHR, or UKHSA.). S.B. acknowledges support from the Novo Nordisk Foundation via The Novo Nordisk Young Investigator Award (NNF20OC0059309).  SB acknowledges the Danish National Research Foundation (DNRF160) through the chair grant which also supports JCS.  S.B. acknowledges support from The Eric and Wendy Schmidt Fund For Strategic Innovation via the Schmidt Polymath Award (G-22-63345) which also supports FMA.

\section*{Author Contributions}

All authors contributed to the Formal Analysis, Methodology, Software, Writing - Original Draft Preparation and Writing - Review \& Editing. JCS conducted the Conceptualization, Validation and Visualization. SB provided Supervision.

\section*{Conflicts of Interest}
The authors declare no conflicts of interest. 
\printbibliography

\end{document}